# Phosphorene FETs – Promising Transistors Based on a few Layers of Phosphorus Atoms

Kuanchen Xiong, Xi Luo, and James C. M. Hwang

Department of Electrical and Computer Engineering
Lehigh University, Bethlehem, PA 18015-3181, USA
Email: jh00@lehigh.edu

*Abstract* — This paper reviews the emergence and progress of phosphorene FETs, all within about a year. In such a short time, back-gated FETs evolved into top-gated FETs, gate length was reduced to the sub-micron range, passivation by high-*k* dielectrics or hexagonal boron nitride was demonstrated with temporal, thermal and mechanical stability, ohmic contact was achieved down to cryogenic temperatures, and cutoff frequencies was pushed above 10 GHz. These and other attractive characteristics of phosphorene promise the phosphorene FET to be a viable candidate for current-generation flexible electronics as well as future-generation ultra-thin-body low-power-consumption high-speed and high-frequency transistors.

TABLE I
UNPASSIVATED PHOSPHORENE FETS

| Organization | Carrier Mobility ($cm^2/Vs$) | Current Capacity (mA/mm) | On/Off Ratio | Subthreshold Slope (V/decade) | Ref. |
|---|---|---|---|---|---|
| Purdue | $10^2$ | $10^2$ | $10^4$ | 1 | [9] |
| Singapore | $10^2$ | | $10^3$ | $10^2$ | [10] |
| Fudan | $10^3$ | 1 | $10^5$ | 10 | [11] |
| Delft | $10^2$ | $10^{-1}$ | $10^3$ | 1 | [12] |
| Yale | $10^2$ | $10^2$ | $10^5$ | 10 | [13] |
| Argonne | $10^2$ | $10^2$ | $10^5$ | 1 | [14] |

## I. INTRODUCTION

Besides graphene, phosphorene is the only other elemental two-dimensional (2D) atomic-layer material that can be mechanically exfoliated, because black phosphorus has [1] a layered honeycomb structure with orthorhombic symmetry held together by van der Waals forces similar to graphite. Black phosphorus is more stable than other allotropes such as red and white phosphorus. It was first synthesized [2] from red phosphorus a century ago under high pressure. It was further explored mainly in Japan with successful synthesis of centimeter-sized crystals [3]. It was found to be naturally *p*-type with a direct bandgap of 0.3 eV [4], which was predicted [5] to increase to approximately 2 eV in a monolayer while remaining direct. In black phosphorus, the electron mobility was measured to be as high as 15,000 $cm^2/Vs$, whereas the hole mobility was as high as 50,000 $cm^2/Vs$ [6]. Since the 1980s [7], there had been little research on black phosphorus except for lithium batteries and mineralization synthesis.

Since 2013, following the explosive interest in graphene and 2D transition-metal dichalcogenides (TMDs), there has been a renewed interest in phosphorene as a new 2D material for electronic applications. This is mainly because phosphorene is unique among all 2D materials by having both an intrinsic and sizable bandgap (unlike graphene) and a high carrier mobility (unlike most TMDs). This paper reviews the recent progress with emphasis on experimental results obtained on phosphorene field-effect transistors (FETs). In this paper, for convention and convenience, the term "phosphorene" is used for trivalent phosphorus and is synonymous to "few-layer black phosphorus" or "2D phosphane" [8].

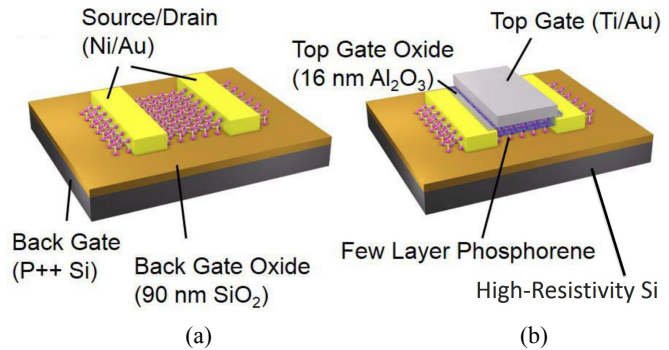

Fig. 1. (a) Back- and (b) top-gated phosphorene FETs [15].

## II. UNPASSIVATED PHOSPHORENE FETS

Table I lists chronologically the pioneering publications on phosphorene FETs [9]−[14], which experimentally demonstrated promising performance characteristics despite crude construction. As shown in Fig. 1(a), these FETs were typically constructed on approximately 10-nm-thick phosphorene exfoliated and transferred onto a degenerately doped silicon substrate coated with approximately 100 nm of $SiO_2$. The silicon substrate served as the back gate while the $SiO_2$ layer served as the gate insulator. The top surface of the phosphorene channel was either exposed or protected only by polymer. The channel length and width were on the order of 1 μm. Measured on these FETs at room temperature, typically, the carrier mobility was on the order of $10^2$ $cm^2/Vs$, the current capacity was on the order of $10^2$ mA/mm, the on/off ratio was on the order of $10^4$, and the subthreshold slope was on the order of 1 V/decade, which corresponded to an

## TABLE II
### PASSIVATED PHOSPHORENE FETs

| University | Gate | Passivation | Ambient Stability | Thermal Stability | Ref. |
|---|---|---|---|---|---|
| KIST | Back | $Al_2O_3$ | 2 mo | | [19] |
| Northwestern | Back | $AlO_x$ | 2 wk | | [20] |
| Lehigh | Top | $Al_2O_3$ | 3 mo | -50 °C−150 °C | [21] |
| UT Austin | Back | $Al_2O_3$ | 3 mo | | [22] |
| UC Riverside | Back | hBN | 2 wk | | [23] |
| Singapore | Top | hBN | 2 mo | -263 °C−25 °C | [24] |
| HKUST | Back | hBN | 1 wk | | [25] |
| Manchester | Back | hBN | Several mo | | [26] |

## TABLE III
### HIGH-FREQUENCY PHOSPHORENE FETs

| University | Gate Length (μm) | Passivation | Substrate | Cut-off Frequencies (GHz) | Ref. |
|---|---|---|---|---|---|
| Lehigh | 0.7 | $Al_2O_3$ | SI GaAs | 1 | [21] |
| USC | 0.3 | $HfO_2$ | HR Si | 10 | [28] |

interface state density $D_{IT}$ on the order of $10^{13}$ cm$^{-2}$ [9]. Thus, although the carrier mobility, current capacity, and on/off ratio appeared adequate, further improvement in the subthreshold slope was needed. Fig. 1(b) shows a more recent construction of a phosphorene FET with a top gate and a high-resistivity substrate for high-speed and high-frequency applications [15]. The top phosphorene surface was passivated with $Al_2O_3$ as discussed in the following.

## III. PASSIVATED PHOSPHORENE FETs

Due to a permanent out-of-plane dipole moment [16], the phosphorene surface is hydrophilic and oxidizable especially when illuminated [17]. It was found [10], [18] that within approximately an hour of ambient exposure, the phosphorene surface became measurably rougher by atomic force microscopy and water droplets were observable by optical microscopy. Thus, although phosphorene is sufficiently stable to allow FETs to be quickly fabricated and characterized as listed in Table I, effective surface passivation is paramount.

To passivate the phosphorene surface, atomic-layer deposited aluminum oxide [19]−[22] and exfoliated hexagonal boron nitride (hBN) [23]−[26] have been used to demonstrate phosphorene FETs that were stable under ambient conditions up to several months as listed in Table II. Thermal stability of some passivated phosphorene FETs were demonstrated [21], [24] from −263 °C to 150 °C, which should be suitable for most device applications especially with a top gate. Additionally, the high- and low-temperature stabilities implied that the source and drain contacts were ohmic and the charge carriers did not freeze out, respectively. The mechanical stability of $Al_2O_3$-passivated phosphorene FETs was demonstrated with 5000 2% strain cycles [27]. The hBN sandwich (Fig. 2) appeared to be of very high quality when carefully assembled [28], which could be annealed up to 500 °C (black phosphorus would transform to red phosphorus at 550 °C) for stable carrier mobility on the order of $10^3$ cm$^2$/Vs and on/off ratio on the order of $10^5$ [25]. The subthreshold slope could also be improved to the order of 0.1 V/decade [23]. In addition to $Al_2O_3$ and hBN, passivation by hafnium oxide appeared [28], [29] to enhance carrier mobility, although stability was not yet reported.

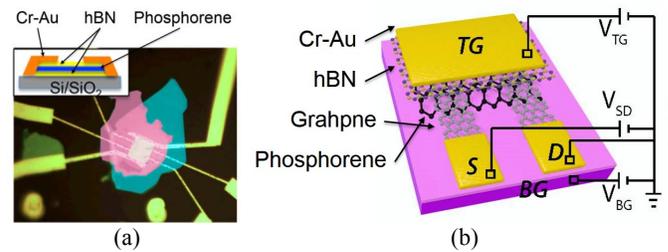

Fig. 2. (a) CrAu [23] and (b) graphene [24] edge contacts to phosphorene.

## IV. HIGH-FREQUENCY PHOSPHORENE FETs

With proper passivation, top-gated phosphorene FETs on a high-resistivity substrate can be expected to perform at radio and microwave frequencies. Table III compares the only two cases reported [21], [28] to date on high-frequency characteristics of phosphorene FETs. The order-of-magnitude difference in cut-off frequencies such as $f_T$ and $f_{MAX}$ between the two cases cannot be explained by the difference in gate length alone, and is probably enhanced by the differences in the passivation and substrate. In both cases, $f_{MAX}$ is greater than $f_T$, which is opposite to that of graphene FETs making phosphorene FETs more suitable for large-signal operation.

## V. CONTACT AND MOBILITY IMPROVEMENTS

Similar to 2D TMD FETs, most phosphorene FETs are made with top contacts (Fig. 1) for the source and drain and can suffer from Schottky barriers [30]. In contrast to 2D TMDs, edge contacts (Fig. 2) to phosphorene appear to be relatively straightforward with graphene [24] or conventional contact metals such as CrAu [23] and can remain ohmic to below 10 K. This may be due to the fact that at 10 nm the phosphorene bandgap approaches the low value of 0.3 eV of black phosphorus. Nevertheless, the achieved contact resistivity is on the order of 1 Ω·cm, which needs to be further reduced. Actually, the contact resistivity is tunable by the back gate, so it is difficult to characterize the contact with a single resistivity.

The carrier mobility measured at room temperature on the "old" (barely one-year old) phosphorene FETs listed in Table I was mostly on the order of $10^2$ cm$^2$/Vs, which was not much higher than that of 2D TMD FETs. However, with proper passivation and atomically flat substrate, there has been [25] report of room-temperature field-effect mobility on the order of $10^3$ cm$^2$/Vs, which is approaching that of black phosphorus. The high quality of the phosphorene is also evidenced by quantum oscillations such as Shubnikov de Haas magneto-oscillations [23], [25], [26], [31], [32]. Although it is encouraging that the confinement of thin phosphorene alone is sufficient to allow quantum oscillations to be observed, heterostructures between phosphorene and other 2D materials may allow more quantum phenomena to be observed and exploited. To this end, van der Waals diodes between *p*-P and *n*-MoS$_2$ [33] and *n*-SnSe$_2$ [34] with staggered (type II) and broken (type III) bandgap alignments, respectively, have been demonstrated. They exhibited interesting tunable rectifying characteristics and negative differential resistance, which could lead to phosphorene tunneling FETs eventually.

## VI. CONCLUSION

The attractive characteristics of phosphorene FETs demonstrated to date, such as current capacity, on/off ratio, cut-off frequencies, temporal, thermal and mechanical stabilities indicate that they are viable candidates for flexible electronics. However, for high-speed and high-frequency applications, they need to be further improved especially in reducing the interface state density to the $10^{11}$-cm$^{-2}$ range and the gate length to the 10-nm range. For such an ultrathin-body low-power consumption nano-FET, not only the gate stack, but also the entire stack from the top gate, top insulator, phosphorene channel, back insulator, back gate, to substrate needs to be optimized together, in addition to optimization of ohmic contacts and proper scaling of all other dimensions. This daunting task may be accomplished in much shorter time than the development of Si or III-V FETs, judging from the extremely rapid development of phosphorene FETs to date. Ultimately, the same as for all 2D devices, uniform, reproducible and large-area growth or synthesis of phosphorene will be required for low-cost high-yield manufacture.

## ACKNOWLEDGEMENT

This work was supported in part by the U.S. Department of Defense, Office of Naval Research, Arlington, VA, USA, under Grant N00014-14-1-0653.

## REFERENCES

[1] R. Hultgren, N. S. Gingrich, and B. E. Warren, "The atomic distribution in red and black phosphorus and the crystal structure of black phosphorus," *J. Chem. Phys.,* vol. 3, pp. 351–355, Jun. 1935.

[2] P. M. Bridgman, "Two new modifications of phosphorus," *J. Am. Chem. Soc.,* vol. 36, pp. 1344–1363, Jul. 1914.

[3] S. Endo *et al.*, "Growth of large single crystals of black phosphorus under high pressure," *Jpn. J. Appl. Phys.,* vol. 21, pp. L482–L484, 1982.

[4] R. W. Keys, "The electrical properties of black phosphorus," *Phys. Rev.,* vol. 92, no. 3, pp. 580–584, Nov. 1953.

[5] Y. Takao, H. Asahina, and A. Morita, "Electronic structure of black phosphorus in tight binding approach," *J. Phys. Soc. Jpn.,* vol. 50, no. 10, pp. 3362–3369, Oct. 1981.

[6] Y. Akahama, S. Endo, and S. Narita, "Electrical properties of black phosphorus single crystals," *J. Phys. Soc. Jpn.,* vol. 52, no. 6, pp. 2148–2155, Jun. 1983.

[7] A. Morita, "Semiconducting black phosphorus," *Appl. Phys. A Solids Surf.,* vol. 39, no. 4, pp. 227–242, Apr. 1986.

[8] *IUPAC Compendium of Chemical Terminology,* 2nd ed., Blackwell Scientific Publications, Oxford, UK, 1997. Available: http://goldbook.iupac.org.

[9] H. Liu *et al.*, "Phosphorene: An unexplored 2D semiconductor with a high hole mobility," *ACS Nano,* vol. 8, no. 4, pp. 4033–4041, Mar. 2014.

[10] S. P. Koenig *et al.*, "Electric field effect in ultrathin black phosphorus," *Appl. Phys. Lett.,* vol. 104, no. 10, pp. 103106-1–103106-4, Mar. 2014.

[11] L. Li *et al.*, "Black phosphorus field-effect transistors." *Nat. Nano.,* vol. 9, no. 5, pp. 372−377, May 2014.

[12] M. Buscema *et al.*, "Fast and broadband photoresponse of few-layer black phosphorus field-effect transistors." *Nano Lett.,* vol. 14, no. 6, pp. 3347−3352, May 2014.

[13] F. Xia, H. Wang, and Y. Jia, "Rediscovering black phosphorus: A unique anisotropic 2D material for optoelectronics and electronics." *Nat. Commun.,* vol. 5, no. 4458, pp.1−6, July 2014.

[14] S. Das *et al.*, "Tunable transport gap in phosphorene," *Nano Lett.,* vol. 14, no. 10, pp. 5733–5739, Aug. 2014.

[15] H. Liu *et al.*, "The effect of dielectric capping on few-layer phosphorene transistors: Tuning the Schottky barrier heights," *IEEE Electron Device Lett.,* vol. 35, no. 7, pp. 795−797, Jul. 2014.

[16] Y. Du *et al.*, "Ab initio studies on atomic and electronic structures of black phosphorus," *J. Appl. Phys.,* vol. 107, pp. 093718-1−093718-4, May 2010.

[17] J. O. Island *et al*, "Environmental stability of few-layer black phosphorus," *2D Mater.,* vol. 2, no. 1, pp. 011002-1−011002-6, Mar. 2015.

[18] A. Castellanos-Gomez *et al.*, "Isolation and characterization of few-layer black phosphorus," *2D Mater.,* vol. 1, no. 2, pp. 025001-1−025001-19, Jun. 2014.

[19] J. Na *et al.*, "Few-layer black phosphorus field-effect transistors with reduced current fluctuation," *ACS Nano,* vol. 8, no. 11, pp. 11753–11762, Nov. 2014.

[20] J. D. Wood *et al.*, "Effective passivation of exfoliated black phosphorus transistors against ambient degradation," *Nano Lett.,* vol. 14, no. 12, pp. 6964–6970, Dec. 2014.

[21] X. Luo *et al.*, "Temporal and thermal stability of Al$_2$O$_3$-passivated phosphorene MOSFETs," *IEEE Electron Device Lett.,* vol. 35, no. 12, pp. 1314–1316, Dec. 2014.

[22] J. S. Kim *et al.*, "Toward air-stable multilayer phosphorene thin-films and transistors," *Sci. Rep.,* vol. 5, pp. 8989-1−8989-7, Mar. 2015.

[23] N. Gillgren *et al.*, "Gate tunable quantum oscillations in air-stable and high mobility few-layer phosphorene heterostructures," *2D Mater.,* vol. 2, no. 1, pp. 011001-1−011001-7, Mar. 2015.

[24] A. Avsar *et al.*, "Electrical characterization of fully encapsulated ultra thin black phosphorus-based heterostructures with graphene contacts." Available: http://arxiv.org/abs/1412.1191.

[25] X. Chen *et al.*, "High quality sandwiched black phosphorus heterostructure and its quantum oscillations." Available: http://arxiv.org/abs/1412.1357.

[26] Y. Cao *et al.*, "Quality heterostructures from two dimensional crystals unstable in air by their assembly in inert atmosphere." Available: http://arxiv.org/abs/1502.03755.

[27] W. Zhu *et al.*, "Flexible black phosphorus ambipolar transistors circuits and AM demodulator," *Nano Lett.,* vol. 15, no. 3, pp. 1883–1890, Mar. 2015.

[28] H. Wang *et al.*, "Black phosphorus radio-frequency transistors," *Nano Lett.,* vol. 14, no. 11, pp. 6424–6429, Nov. 2014.

[29] N. Haratipour, M. C. Robbins, and S. J. Koester, "Black phosphorus p-MOSFETs with 7-nm HfO$_2$ gate dielectric and low contact resistance," *IEEE Electron Device Lett.,* vol. 36, no. 4, pp. 411–413, Apr. 2015.

[30] Y. Du *et al.*, "Device perspective for black phosphorus field-effect transistors: Contact resistance, ambipolar behavior, and scaling," *ACS Nano,* vol. 8, no. 10, pp. 10035–10042, Oct. 2014.

[31] L. Li *et al.*, "Quantum oscillations in black phosphorus two-dimensional electron gas." Available: http://arxiv.org/abs/1411.6572.

[32] L. Li *et al.*, "Quantum Hall effect in black phosphorus two-dimensional electron gas." Available: http://arxiv.org/abs/1504.07155.

[33] Y. Deng *et al.*, "Black phosphorus-monolayer MoS$_2$ van der Waals heterojunction p-n diode," *ACS Nano,* vol. 8, no. 8, pp. 8292–8299, Aug. 2014.

[34] R. Yan *et al.*, "Esaki diodes in van der Waals heterojunctions with broken-gap energy band alignment." Available: http://arxiv.org/abs/1504.02810.